\documentclass[aps,amsfonts,pra,twocolumn,showpacs]{revtex4-1}

\usepackage{pdfpages} 
\makeatletter
\AtBeginDocument{\let\LS@rot\@undefined}
\makeatother

\usepackage{subfigure}
\usepackage{textcomp}
\usepackage{graphicx}
\usepackage{amssymb}
\usepackage{amsmath}
\usepackage{bm}
\usepackage{amsmath, amsthm, amssymb}
\usepackage{dsfont}
\usepackage[colorlinks=true,linkcolor=blue,urlcolor=blue,citecolor=blue]{hyperref}
\usepackage{multirow}
\usepackage{url}

\def\beq{\begin{equation}}
\def\eeq{\end{equation}}
\def\bea{\begin{eqnarray}}
\def\eea{\end{eqnarray}}

\begin{document}

\title{Anomalous relaxation and the high-temperature structure factor of XXZ spin chains}

\author{Sarang Gopalakrishnan$^{1}$, Romain Vasseur$^{2}$, and Brayden Ware$^{2,3}$}
\affiliation{$^1$ Department of Physics and Astronomy, CUNY College of Staten Island, Staten Island, NY 10314;  Physics Program and Initiative for the Theoretical Sciences, The Graduate Center, CUNY, New York, NY 10016, USA}
\affiliation{$^2$ Department of Physics, University of Massachusetts, Amherst, MA 01003, USA}
\affiliation{$^3$ Rudolf Peierls Centre for Theoretical Physics, Clarendon Laboratory, University of Oxford, Oxford OX1 3PU, UK}

\begin{abstract}

We compute the spin structure factor of XXZ spin chains in the Heisenberg and gapped (Ising) regimes in the high-temperature limit for nonzero magnetization, within the framework of generalized hydrodynamics including diffusive corrections. The structure factor shows a hierarchy of timescales in the gapped phase, owing to $s$-spin magnon bound states (``strings'') of various sizes. 
Although short strings move ballistically, long strings move primarily diffusively as a result of their collisions with short strings. The interplay between these effects gives rise to anomalous power-law decay of the spin structure factor, with continuously varying exponents, at any fixed separation in the late-time limit. 
We elucidate the crossover to diffusion (in the gapped phase) and to superdiffusion (at the isotropic point) in the half-filling limit. 
We verify our results via extensive matrix product operator calculations.

\end{abstract}

\maketitle


%
Many experimentally relevant one-dimensional systems are described by approximately integrable models, such as the Hubbard, Heisenberg, and Lieb-Liniger models~\cite{kinoshita, bloch_heisenberg, Rigol:2008kq}. The nonequilibrium dynamics of integrable systems, their failure to thermalize, and their possession of an extensive set of conservation laws, have been explored extensively~\cite{PhysRevLett.110.257203, 2016arXiv160300440I, PhysRevLett.115.157201, PhysRevLett.113.117203}. (In experiments, integrability is approximate, and gives rise to ``prethermal'' intermediate-time regimes of effectively integrable dynamics~\cite{gring, 2016arXiv160309385L, tang2018}.)
%
%
Integrable systems support stable, ballistically propagating quasiparticles even at high temperature. In the simplest cases (e.g., free fermions), these particles carry the same quantum numbers as the microscopic degrees of freedom, and move with a velocity set by the band structure. In \emph{interacting} integrable models, however, each quasiparticle is dressed by all the others~\cite{Takahashi}. This dressing can lead to remarkable dynamical effects, for instance in the ``gapped'' phase of the XXZ model considered here: here, even though quasiparticles move ballistically, finite-temperature spin transport is diffusive in the absence of an external field~\cite{PhysRevLett.78.943,PhysRevB.57.8307,PhysRevLett.82.1764, PhysRevLett.95.187201, sirker:2010,PhysRevB.89.075139, lzp, piroli2017, bertini2018low, dbd1, ghkv, dbd2, gv_superdiffusion, denardis_superdiffusion}. 

Recently, a coarse-grained approach to integrable dynamics has been developed; this approach is termed ``generalized hydrodynamics'' (GHD)~\cite{Doyon, Fagotti}, see also~\cite{SciPostPhys.2.2.014,GHDII, PhysRevLett.119.020602, BBH, solitongases, piroli2017, PhysRevB.96.081118,PhysRevB.97.081111,alba2017entanglement,PhysRevB.96.020403, bertini2018low, BBH0,1751-8121-50-43-435203, PhysRevLett.119.195301, 2017arXiv171100873C, PhysRevLett.120.164101,doyon2018exact,PhysRevLett.122.090601,2019arXiv190300467A}. A core insight of GHD is that an integrable system can be mapped to an appropriate classical soliton gas~\cite{solitongases}. Assuming the system is initially in local equilibrium, the velocities of these solitons can be computed using the thermodynamic Bethe ansatz~\cite{PhysRevLett.113.187203,Doyon, Fagotti}, which is much more tractable than exactly simulating the full dynamics. In the initial formulation of GHD, the dressing of quasiparticles by interactions was treated at the ``Euler'' level, yielding purely ballistic hydrodynamics; recently, adding Gaussian fluctuations on top of this treatment was shown to give diffusive corrections to hydrodynamics~\cite{dbd1, ghkv, dbd2,gv_superdiffusion}. In the generic case, diffusive corrections occur on top of ballistic transport; however, in many situations the ballistic term is absent, and transport is dominated by normal or anomalous diffusion. 

\begin{figure}[bt]
\begin{center}
\includegraphics[width = 0.47\textwidth]{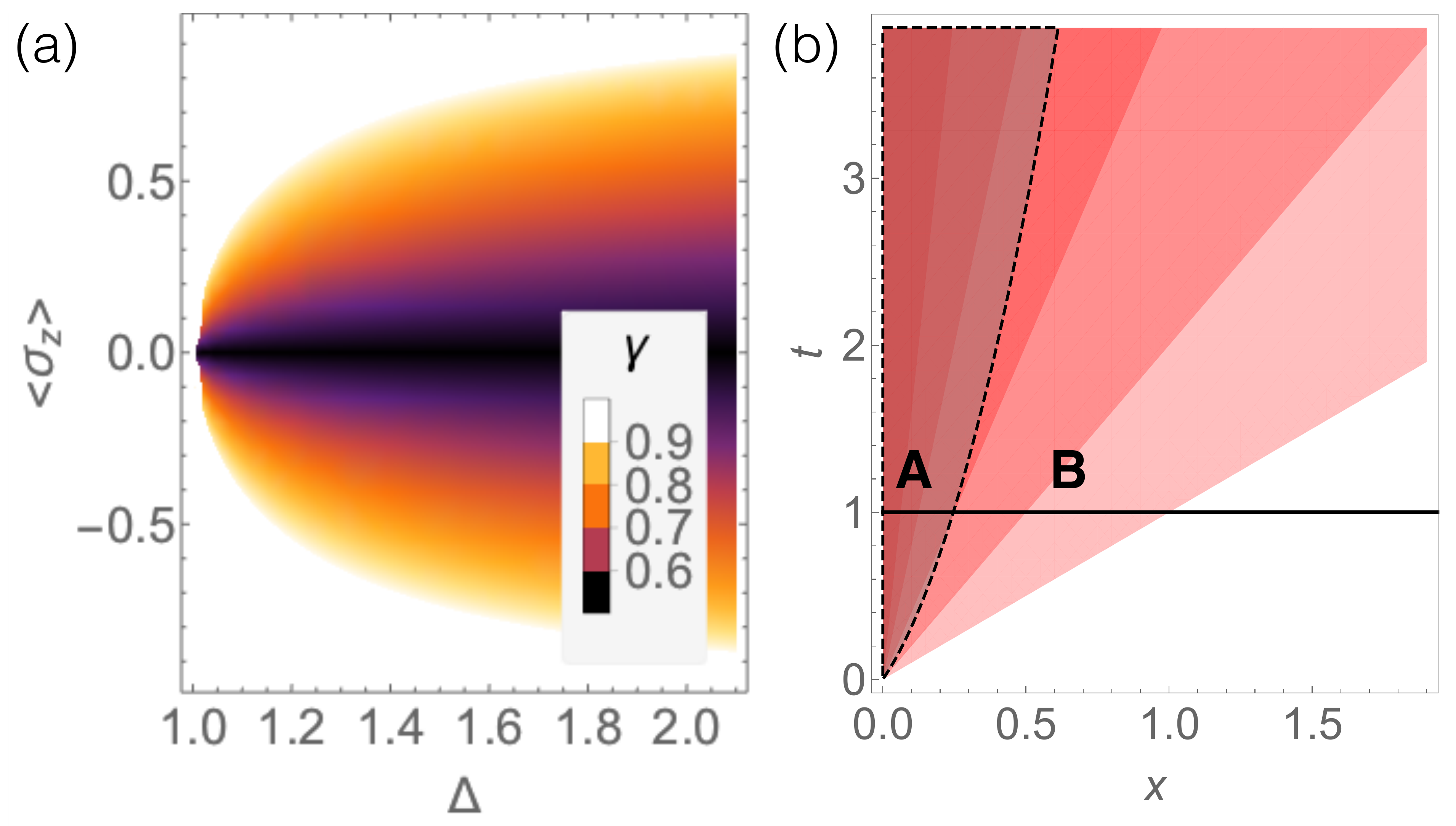}
\caption{{\bf Return probability.}  (a)~Exponent of the return probability, $C(0,t) \sim t^{-\gamma}$, as a function of the filling and the anisotropy. This result applies for any fixed $x$ as $t \rightarrow \infty$. (b)~Mechanism for anomalous local relaxation: the velocity of an $s$-string in the easy-axis XXZ model decreases exponentially with $s$. Light strings in region B spread out ballistically; heavy strings in region A move diffusively because of collisions with light strings. As time passes, more strings become ``light'' in that their motion is chiefly ballistic.}
\label{fig1}
\end{center}
\end{figure}

In this work, we show that even when ballistic transport is present, certain aspects of the structure factor exhibit anomalous exponents. We focus on the XXZ spin chain:
\beq\label{xxz}
H =  \sum\nolimits_i \left[ S^x_i S^x_{i+1} + S^y_i S^y_{i+1} + \Delta S^z_i S^z_{i+1} \right],
\eeq
where $S^\alpha=\frac{\sigma^\alpha}{2}$ are spin-$\frac{1}{2}$ operators. We are concerned with the case of easy-axis anisotropy $|\Delta| \geq 1$ at infinite temperature (so the sign of $\Delta$ is irrelevant). We define $\eta \equiv \cosh^{-1} \Delta$. The model has a conserved magnetization, $\sigma^z_{\rm tot} = \sum\nolimits_i \sigma^z_i$; we denote the associated magnetization density as $h = \tanh \mu$, corresponding to a filling $f=(1+h)/2$. At half-filling ($h = 0$), ballistic transport is absent because the propagating quasiparticles carry no spin. One can easily see this for magnons in the ferromagnetic phase at low but nonzero temperatures: a magnon propagates as a $\downarrow$ spin through a domain of $\uparrow$ spins, then continues as an $\uparrow$ spin through a domain of $\downarrow$ spins, etc., so on average it does not carry magnetization. This result, which holds generally, was first noted in the low-temperature limit~\cite{PhysRevB.57.8307,PhysRevLett.95.187201}, and has recently been incorporated into the GHD framework. 
Since ballistic transport is absent, the dominant transport mechanism is diffusive. The diffusion constant $D$ has been rigorously bounded~\cite{idmp} and computed using GHD~\cite{dbd2, gv_superdiffusion, denardis_superdiffusion}; $D$ diverges in the isotropic ($\Delta = 1$) limit, at which superdiffusion takes place, with a time-dependent diffusion constant $D(t) \sim t^{1/3}$; wavepackets appear to spread with a non-Gaussian front that corresponds to the Kardar-Parisi-Zhang (KPZ) universality class~\cite{lzp, ljubotina2019kardar, kpz, quastel}. Away from half-filling, the $\uparrow$ and $\downarrow$ domains do not precisely cancel, so magnons do carry magnetization and ballistic spin transport is present. Beyond these results, the behavior of the dynamical spin structure factor and the optical conductivity in this high-temperature limit are still poorly understood. (There has been extensive work on these quantities at zero temperature~\cite{PhysRevLett.95.077201, PhysRevLett.96.257202, pereira2007dynamical, sirker2008boson, PhysRevLett.106.157205} and in the low-energy field theory regime~\cite{ PhysRevLett.78.943,PhysRevLett.95.187201,PhysRevB.57.8307,sirker:2010,PhysRevB.83.035115}, but as we will see the physics is qualitatively different at high temperature.)

The present work addresses these issues, computing the spin structure factor within GHD. We focus on the connected correlation function $C(x,t) \equiv \langle S^z_{i + x}  (t)S^z_{i} (0) \rangle^c$ evaluated at infinite temperature with chemical potential $\mu$; everything we discuss will involve large $x, t$ but arbitrary ratios $x/t$. 
We find that, even away from half filling, the local behavior of autocorrelators (i.e., for $x/t \ll 1$, corresponding to the return probability) evolves with continuously varying exponents that depend on $\Delta$ and $h$ (Fig.~\ref{fig1}). There is a phase transition in the $(\Delta, h)$ plane, between ballistic (i.e., $1/t$) and sub-ballistic (i.e., $1/t^\gamma$ for $1/2 < \gamma < 1$) asymptotic behavior. We compute the exponent $\gamma$ as a function of $(h, \Delta)$, and show that for $\Delta > 1$ it universally approaches $1/2$ as $h \rightarrow 0$, recovering (and shedding light on) diffusion at half-filling. This coexistence of ballistic and anomalous behavior was recently demonstrated~\cite{agrawal2019} for disordered integrable spin chains~\cite{evg2018}; here we show that the same effect occurs in clean strongly interacting systems. 
At the Heisenberg point, the phase boundary in the $(\Delta, h)$ plane intersects the ballistic-diffusive phase boundary at $h = 0$, and in this sense the isotropic Heisenberg point at $h = 0$ is a dynamical multicritical point. We write down a scaling form for the structure factor as one approaches this critical point at finite $h$.

\emph{Low-filling limit}.---Our results have an elementary interpretation in the limit in which $f \ll 1$. Here, $f \sim e^{2\mu}$ with $\mu \to -\infty$. Nevertheless the system is still at infinite temperature. Further, for the present discussion we take $\Delta \gg 1$. Under these conditions we can calculate the structure factor by elementary methods; we only invoke integrability to claim that quasiparticles are in fact stable. In this limit, the quasiparticles are essentially ``bare'': an $s$-string is a sequence of $s$ $\uparrow$ spins on top of a $\downarrow$ background. Since an $s$-string can only move at $s$th order in perturbation theory, its velocity is $v_s \sim \Delta^{1 - s}$. Neglecting dressing, the $s$-strings have free-particle dispersions of the form
\beq\label{sstring}
\epsilon_s(q) = k_s \Delta^{1-s} \sin(2q),
\eeq
where $k_s$ are constants of order unity. In our discussion of this limit, we take $f \sim {\rm e}^{2 \mu} \to 0$ and  $\Delta \sim {\rm e}^\eta \to \infty$, but allow the ratio $\left| \mu \right|/  \eta$ to be of order unity. In the following, we will use GHD to generalize our results to arbitrary filling and $\Delta >1$. 

If we ignore diffusive corrections, the model is in effect a gas of free $s$-strings, which occur with probability $f^s$. All dressing effects are suppressed by factors of $f$ with no compensating factors of $1/\Delta$, so we neglect them. Then a single-particle calculation yields the structure factor, as follows:
\beq\label{euler}
C(x,t) \simeq \sum\nolimits_{s \geq 1} s^2 f^s [J_x(k_s \Delta^{1-s} t)]^2,
\eeq
where $J_x$ denote Bessel functions of the first kind~\cite{fukuhara2013quantum}. 
A nontrivial contribution arises if a string beginning at $(0,0)$ has propagated to $(x,t)$.
To explore the asymptotics of Eq.~\eqref{euler} we approximate the Bessel function as a step function and ignore irrelevant prefactors, $[J_x(k_s \Delta^{1-s} t)]^2 \sim \Theta(x - \Delta^{1-s} t) \Delta^{s-1}/t$. Fixing a point $x$, and counting only those $s$-strings that have reached $x$ by the time $t$, we get 
\beq\label{approxeuler}
C(x,t) \approx \sum\nolimits_{s = 1}^{s_*} s^2 \frac{f}{t} \left( f\Delta \right)^{s-1},
\eeq
where $s_* = 1 + \log(t/x) / \log \Delta$. There are two cases. When $f \Delta < 1$, higher-order strings are too rare to contribute to the correlation function, which is dominated by the $1/t$ tail of the 1-strings. When $f \Delta > 1$, the dominant strings at position $x$ are the heaviest strings that have made it there; the sum in Eq.~\eqref{approxeuler} is given by the term of order $s_*$. This then gives the asymptotics
\beq\label{asymp1}
C(x,t) \! \sim \! \frac{f}{t} \! \left( \frac{t}{x} \right)^{1 - \frac{|\log f|}{\log \Delta}} \!\! \log^2\!\left( \frac{t}{x} \right) \sim t^{- \frac{2 \left| \mu \right|}{ \eta}} \log^2\! t , 
\eeq
for  $2| \mu| < \eta$. 
The exponent $\gamma =\frac{2 \left| \mu \right|}{\eta}$ in Eq.~\eqref{asymp1} goes to unity as $\eta \gg | \mu |$, suggesting subdiffusion ($\gamma < \frac{1}{2}$) as $\Delta \rightarrow \infty$ at the Eulerian level. This asymptotic decay will occur through a series of jumps~\cite{piroli2017}; we assume here that one coarse-grains over long enough time windows to average out these features.

\emph{Diffusive corrections}.---The asymptotics~\eqref{asymp1} arises because---at the Eulerian level---long strings are assumed to be effectively stationary for exponentially long times. At finite $f$ this is not, in fact, the case: when a $q$-string and an $s > q$-string collide, the $s$-string picks up a displacement of $2q$ sites~\cite{quantumbowling, ganahl_caux}. Thus, all strings undergo subleading diffusive motion. In the low-density limit it suffices to consider the diffusion of $s$-strings due to collisions with 1-strings, so the diffusion constant scales as $ D \sim f$. Because the model is integrable this diffusion takes place \emph{in addition to} the ballistic motion of $s$-strings. At intermediate times, an $s$-string moves diffusively; however, there is a crossover to ballistic motion at times such that $\sqrt{f t} < \Delta^{1-s} t$, i.e., for strings satisfying
\beq\label{tn}
s \agt s_0(t) \sim 1 + \frac{|\log (ft)|}{2 \log \Delta}.
\eeq
When $f \Delta < 1$, heavy strings are sparse so it does not matter whether they diffuse. 
When $f \Delta > 1$, Euler-scale results remain valid at distances $x \agt x_0 \equiv \sqrt{f t}$, but the behavior at distances $x \alt x_0$ is qualitatively modified. Rather than staying immobile, strings with $s > s_0(t)$ move diffusively. Thus the autocorrelator decays as
\beq
C(x \alt x_0,t) \sim t^{-\frac{1}{2} - \frac{\left| \mu \right| }{\eta}} \log^2 t , \quad  2 \left| \mu  \right|<  \eta.
\label{eqAnomalousRelaxation}
\eeq
The slowest that $C(x,t)$ can decay is as $1/\sqrt{t}$, {\it i.e.}, a subdiffusive decay of the return probability does not occur in this model. Instead, the generic behavior is an anomalous decay with a continuously varying power law between $\frac{1}{2}$ and $1$.

\emph{GHD approach at generic filling}.---
The above argument is elementary but is restricted to the limit of large $\Delta$ and low filling. We now show that our main conclusion -- the anomalous decay~\eqref{eqAnomalousRelaxation} of the local autocorrelation function -- holds generally for all $h$ whenever $\Delta > 1$. In the general case, spin transport can still be understood in terms of a hierarchy of strings, but their interactions are now important and their velocity and effective charge are dressed by the collisions with other quasiparticles. These issues can be addressed using GHD: since we are dealing with a linear response problem, we take advantage of the fact that the quasiparticles are in local thermal equilibrium, and evaluate the dressed quasiparticle dispersion and quasiparticle distribution function using data from the thermodynamic Bethe ansatz solution.
 Then the hydrodynamic expression for the structure factor takes the form~\cite{GHDII,PhysRevB.96.081118}:
\beq
C(x,t)   =  \sum_{s = 1}^\infty \!\int du\, \rho^{\rm tot}_s(u) \theta_s (1-\theta_s) (m^{\mathrm{dr}}_s)^2 \varphi_t[x - v_s(u) t], \label{eqGHDSzSz}
\eeq
where $u$ parameterizes the rapidity of a quasiparticle; $m^{\mathrm{dr}}_s$ is the dressed magnetization of string $s$, $\rho^{\rm tot}(u)$ is the available density of states, $\theta_s$ is its occupation number (Fermi factor), and $v_s$ is its effective velocity. These quantities have closed-form expressions for generic $\mu$ at infinite temperature~\cite{Takahashi,idmp,suppmat}. Finally, the function $\varphi_t (\zeta)$ is the propagator of a string with quantum numbers $(s, u)$ from $(0,0)$ to $(x,t)$. At the Euler level this propagator would simply be a Dirac delta function. In principle the full form of $C(x,t)$, including diffusive corrections and possible nonlinearities, could be ascertained from flea-gas simulations~\cite{solitongases}. 
Here we are interested in the asymptotic behavior of this quantity. We therefore include the dominant ``diagonal'' diffusive corrections by broadening $\varphi_t(\zeta)$ to a Gaussian with variance $2 D_{s}(\eta,u) t$. The diagonal quasiparticle diffusion constant $D_{s}(\eta,u)$ was computed in Refs.~\cite{dbd1,ghkv}, and can be evaluated numerically. We can check explicitly that our hydrodynamic form for the structure factor~\eqref{eqGHDSzSz} is consistent with the exact sum rule 
\beq
\int_{-\infty}^{\infty} dx\, C(x,t) = \frac{1}{4} (1 - \tanh^2 \mu), \label{eqSumRule}
\eeq
since the function $\varphi_t(\zeta)$ is normalized to unity~\cite{suppmat}.

\emph{Anomalous local relaxation}.---Equipped with this GHD expression, we first consider local relaxation, i.e., $C(x,t)$ at fixed large $x$ when $t \rightarrow \infty$. There are two contributions at time $t$, from light strings (whose motion is primarily ballistic) and from heavy strings (which undergo Brownian motion from collisions with light strings). Regardless of $\mu$, the velocities of very heavy strings scale as $v_s \simeq e^{-\eta s}$; also, at infinite temperature, their densities scale as $\rho_s(u)= \rho^{\rm tot}_s(u) \theta_s \sim e^{-2|\mu| s}$~\cite{dbd2}. 
The dressed magnetization of the heavy strings, meanwhile, is the same as the bare magnetization $m^{\rm dr}_s \simeq s$. 
We see that the asymptotics of $v_s$ and $\rho_s$ are identical to the low-filling limit: for $\eta > 2 \left| \mu \right|$, the return probability 
is dominated by the diffusive strings, $s > s_0(t)$, where $s_0(t) \simeq \frac{1}{2\eta} \log t$. It follows that Eq.~\eqref{eqAnomalousRelaxation} applies for all $\mu$ and $\eta > 1$. (Note, however, that away from the perturbative limit $\eta \neq \log \Delta$.)

\begin{figure*}[!t]
\begin{center}
\includegraphics[width = 0.95\textwidth]{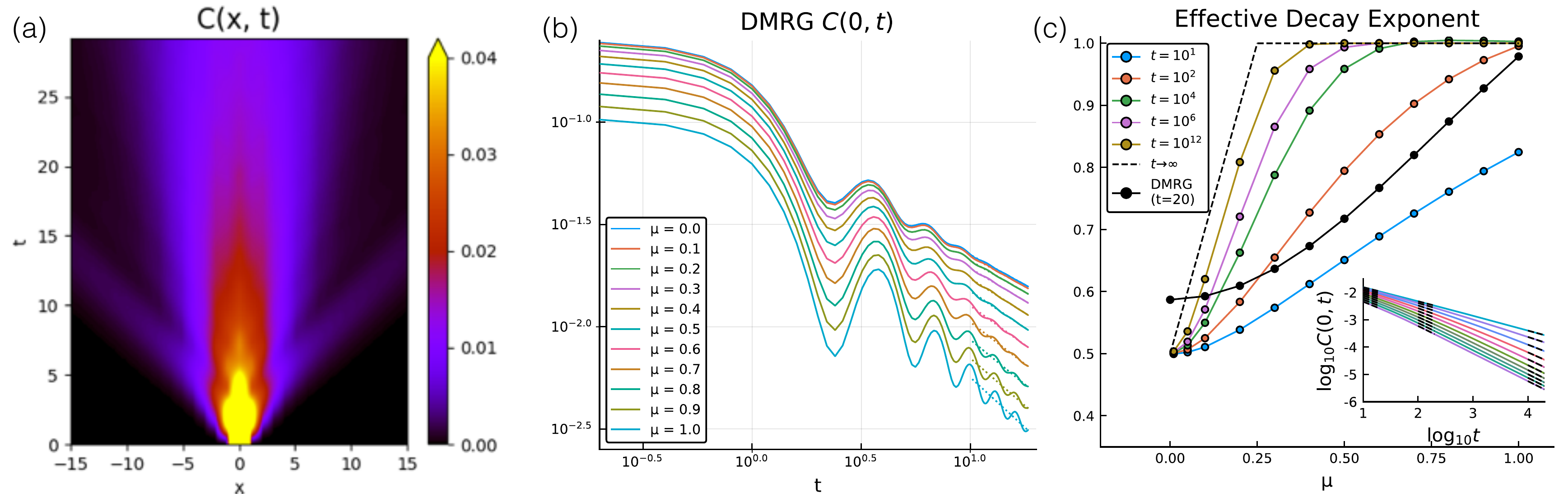}
\caption{{\bf Simulations of the structure factor.} (a)~Spacetime plot of the spectral intensity computed via the MPO method, for $\eta = 1.5, \mu = 0.5$, indicating a ballistically moving peak due to magnons and a slow background to heavier strings. (b)~MPO simulations of the return probability vs. $\mu$ at fixed $\eta = 0.5$; the exponent is initially close to diffusive, then shifts downward with increasing $\mu$. (c)~Comparison of exponents extracted from the MPO simulations to those computed by numerical evaluation of Eq.~\eqref{eqGHDSzSz} using various fitting windows. Evidently the GHD result is quite slow to converge to its late-time asymptotic behavior.}
\label{MainFig}
\end{center}
\end{figure*}

For a fixed $\eta > 0$ and $\mu < \eta/2$, this asymptotic scaling sets in on timescales $t \agt e^{2\eta/|\mu|}$; at shorter times (see below) we expect a smaller apparent exponent, since the dominant strings at those times are not yet exponentially suppressed (Fig.~\ref{MainFig}). Reaching the asymptotic regime on accessible timescales is numerically challenging: to see many heavy strings at short times, we need $\eta \ll 1$, i.e., near the isotropic limit; however, working near the isotropic limit leads to {transient superdiffusion~\cite{lzp,idmp} at short times $t \ll \eta^{-3}$~\cite{gv_superdiffusion}}.

\emph{Properties near half-filling}.---Near half filling, i.e., for $\mu \ll 1$, we can extract more quantitative information about the structure factor. 
Again, we classify strings as light and heavy at time $t$, depending on whether their spread up to time $t$ is primarily ballistic or diffusive. For light strings, diffusive corrections are a subleading effect (except at the front) so we treat light strings at the Euler level. For heavy strings near half filling, the diffusive broadening constant has the closed-form expression
\beq\label{diffcons}
D = \frac{2 \sinh \eta}{9 \pi} \sum_{s=1}^\infty (1+s) \left[\frac{s+2}{\sinh \eta s} - \frac{s}{\sinh \eta (s+2)} \right],
\eeq
which coincides with the spin diffusion constant~\cite{dbd2,gv_superdiffusion, denardis_superdiffusion}. 
This expression is a sum over contributions from $s$-strings, and only strings with $s\eta \alt 1$ contribute. A slight distance $\mu$ away from half filling, the properties of strings with $s \mu \alt 1$ are similar to those at half-filling, while strings with $s \mu \agt 1$ have exponentially suppressed density. When $\mu \ll \eta$ the contributions to Eq.~\eqref{diffcons} that would be modified by finite $\mu$ are already exponentially suppressed in $\eta$,
so heavy strings with $\eta^{-1} \ll s$ diffuse with their $\mu = 0$ diffusion constant~\eqref{diffcons} up to exponentially small corrections.

We now discuss the behavior of $C(x,t)$ near half-filling along rays with $x/t \neq 0$. For nonzero $\mu$, at late times, there are two different types of ballistic strings, depending on the size of $s\mu$. When $s\mu \gg 1$, the ballistic strings behave as in the low-filling limit: their density and velocity are both exponentially suppressed, and we recover Eq.~\eqref{asymp1}. However, for lighter strings with $s\mu \alt 1$, the density is only suppressed algebraically as $\rho_s \simeq 1/s^3$ while the velocity is suppressed exponentially $v_s \simeq e^{-\eta s}$. Therefore, at a fixed position $x$, the largest string that has made it out to $x$ has index $s_* = \log(t/x) / \eta$. The density of such strings is $1/s_*^3$ while each carries a small dressed magnetization $s_*^2 \mu$. Thus, in this regime, we have 
\beq\label{regime3}
C(x,t) \sim \frac{\mu^2}{\eta} \frac{\log(t/x)}{x}, \quad e^{-\eta/|\mu|}  \ll \frac{x}{t} \ll 1,
\eeq
{where the regime of validity of this result is controlled by $\mu$, and is to be understood on a logarithmic scale for $x/t$.} 
Interestingly, the correlator at a fixed position $x$ \emph{grows} logarithmically with time, as heavier strings carrying more magnetization appear at $x$. At longer times, the structure factor decays anomalously as~\eqref{eqAnomalousRelaxation}. Exactly at half filling ($\mu=0$), the structure factor simplifies even further. All strings but the heaviest ones $s \to \infty$ become effectively neutral as $m^{\rm dr}_s \sim s^2 \mu$ goes to 0 for $s \ll \mu^{-1}$, so the structure factor reads
\beq
C(x,t)   =  \frac{1}{4} \varphi_t(x) = \frac{1}{8\sqrt{\pi D(\eta) t}} {\rm e}^{ - \frac{x^2}{4 D(\eta) t}},
\eeq
where we have used the sum rule~\eqref{eqSumRule} at half filling, and $v_{\infty} = 0$. At half-filling, the structure factor is thus given by the heaviest strings~\cite{dbd2}, which are moving purely diffusively because of random collisions with lighter strings, with the spin diffusion constant~\eqref{diffcons}. 

\emph{Matrix product operator calculations}.--- We test these predictions by computing the structure factor $C(x,t)$ in the Heisenberg picture by time evolving $S^z_i$ using matrix product operator (MPO) techniques and the time-dependent density matrix renormalization group (tDMRG)~\cite{whitetdmrg, PhysRevLett.91.147902, schollwoeck,karraschdrude,1367-2630-15-8-083031}. We find that a fixed truncation error $\epsilon =10^{-8}$ is enough to obtain converged results, and we use a fourth order Trotter decomposition with time step $dt = 0.1$. Our calculations are stopped when the bond dimension reaches $\chi \sim 2000$. This approach allows us to compute $C(x,t)$ for any filling (or temperature) from a single calculation of $S^z_i(t)$. Following Ref.~\cite{ljubotina2019kardar} (see also~\cite{VKM,1367-2630-19-3-033027} in the context of the Drude weight), we also compute $C(x,t)$ using a linear response quench setup when the system is initially prepared in a non-equilibrium state with chemical potential $\mu+\delta \mu/2$ in the left half of the system, and $\mu-\delta \mu/2$ in the right half. We then compute the density matrix $\rho(t)$ at time $t$ using MPO methods. Working at fixed truncation error, this approach allows us to reach similar time scales $t \sim 20$ to obtained fully converged results. If we work instead with fixed bond dimension MPOs (with bond dimensions $\chi = 100, 200,300,400$), the quench time-traces deviate from the exact MPO time trace at short times $t \alt 20$ {for the return probability and appear to oscillate about it}, though they give {reasonably converged spatial profiles for $x \neq 0$} out to late times~\cite{suppmat}, as noted in Refs.~\cite{lzp,ljubotina2019kardar}. 

These results are plotted in Fig.~\ref{MainFig}. For {$\mu = 0.5$ and $\eta=1.5$}, the structure factor has a clear ballistic front due to magnons, with a broad diffuse feature behind it, as GHD predicts. The middle panel shows the local autocorrelator (i.e., return probability) as a function of $\mu$ at fixed $\eta = 0.5$. Its behavior is consistent with a continuously varying power law that goes from approximately $1/2$ (in fact closer to $0.6$ {due to the proximity to the isotropic point $\Delta=1$}) at $\mu = 0$ to nearly one at large $\mu$. The numerically extracted exponent $\gamma$ is much smaller than the asymptotic GHD prediction. To understand this discrepancy we have numerically evaluated the return probability using Eq.~\eqref{eqGHDSzSz}. As shown in the right panel of Fig.~\ref{MainFig}, the GHD curves curve downwards at short times, and only converge to their asymptotic slopes at extremely long times, {as anticipated above}.

\emph{Isotropic point}.--- {Finally, we briefly discuss the structure factor at the isotropic point $\eta=0$ ($\Delta=1$). At half-filling, Eq.~\eqref{diffcons} implies that the diffusion constant diverges with the number of strings as $D \sim s$. Spin transport at half-filling is therefore superdiffusive~\cite{lzp,idmp}, with a time-dependent diffusion constant that was argued to scale as $D(t) \sim t^{1/3}$ due to the anomalous behavior of heavy strings~\cite{gv_superdiffusion} (see also~\cite{denardis_superdiffusion}), consistent with numerical results~\cite{lzp,ljubotina2019kardar}. 

By considering the approach to half filling at finite $\mu$ one can retrieve the dynamical exponent $x \sim t^{2/3}$ by a simple intuitive argument. A typical thermal state has Gaussian spatial fluctuations of its magnetization, so the effective local magnetization fluctuates as $1/\sqrt{L}$ over a distance $L$. On short enough length-scales, these fluctuations dominate over the average $\mu$. 
The system averages out these fluctuations and ``realizes'' it is at $\mu \neq 0$ on a length-scale such that $\mu \sim 1/\sqrt{L}$, i.e., the crossover length scales as $L(\mu) \sim 1/\mu^2$. Further, as $\mu \rightarrow 0$, magnetization is primarily transported by the heaviest available strings, for which $s_* \simeq 1/\mu$ and $v_{s_*} \simeq \mu$. The time it takes these strings to travel a distance $L(\mu)$ is given by $t(\mu) \sim 1/\mu^3$. {The diffusion constant of such strings diverges as $D \sim \mu^{-1}$, which also gives the same scaling $t(\mu) \sim 1/\mu^3$. }
 It follows that $L \sim t^{2/3}$. Moreover, the structure factor near half-filling can be written in the scaling form
\beq
C(x,t) = \mu^2 \left[C_{\mathrm{anom.}} (x\mu^2, t\mu^3) + \frac{1}{t} C_{\mathrm{reg.}} (x/t) \right],
\eeq
where the first term comes from strings with $s^* \sim 1/\mu$ and the second from lighter strings. At precisely half-filling the regular part vanishes as $\mu^2$, and only the anomalous part survives. The regimes of $C_{\mathrm{anom.}}(\zeta, \xi)$ are as follows. When both $\zeta, \xi \ll 1$, $C_{\mathrm{anom.}} \sim (t \mu^3)^{-2/3} f(x/t^{2/3})$, where $f$ was numerically found to have the KPZ form~\cite{ljubotina2019kardar}. When $\zeta \gg \xi$ the anomalous part vanishes by causality. The late-time return probability $\xi \gg 1, \zeta \ll \xi$ is dominated by the tail of the heaviest common string, i.e., it goes as $1/(\mu t)$. Putting these together we have
\beq
C(0,t) = t^{-2/3} g(\mu t^{1/3}), \quad g(y) = \left\{ \begin{array}{lr} \mathrm{const.} & y \ll 1 \\ 1/y & y \gg 1 \end{array} \right. 
\eeq
%
Meanwhile, the ballistic, regular part can be calculated following the logic of Eq.~\eqref{regime3}, so $C_{\mathrm{reg.}}(y) \sim 1/y^2$ for $y \ll 1$, implying that
\beq\label{xxxregime3}
C(x/t) \sim \mu^2 t / x^2, \quad \mu \ll (x/t) \ll 1.
\eeq
As $\mu \rightarrow 0$ spatial fluctuations of the magnetization dominate the dynamics. If we imagine dividing the system into a large number of hydrodynamic cells with magnetization $m(x,t)$, each cell will have a fluctuating diffusion constant $D[m] \sim 1/m$ and ballistic spin transport coefficient $j_{\rm ballistic}[m] \sim v[m] m  \sim m^2$ set by its instantaneous magnetization (repeating the argument above with $m$ instead of $\mu$ as the cutoff). Combining these contributions into a hydrodynamic equation for $m$ yields a Burgers equation with a diffusion constant that is singular at low density (see also~\cite{denardis_superdiffusion}). We expect this to be compatible with KPZ scaling~\cite{kpz}: Over a distance $\ell$, Gaussian fluctuations in the initial state lead to $m \sim 1/\sqrt{\ell}$, implying a diffusion constant $D \sim \sqrt{\ell}$ and ballistic velocity $v \sim 1/\sqrt{\ell}$, both implying $t(\ell) \sim \ell^{3/2}$. Moreover, the dominant nonlinearities in the Burgers equation involve anomalous high-density regions, for which the diffusion coefficient is well-behaved, so one might conjecture that the KPZ scaling function is also unaffected, as the numerical evidence~\cite{ljubotina2019kardar} suggests.
However, developing this nonlinear fluctuating hydrodynamics~\cite{spohn_nlfhd} for integrable systems is outside the scope of the present work.

\emph{Discussion}.---In this work we used generalized hydrodynamics and its diffusive corrections to characterize the structure factor of the XXZ model in the easy-axis regime and at the isotropic Heisenberg point. We argued that even at nonzero magnetization, where ballistic transport is present, the local behavior of the autocorrelation function exhibits rich structure due to heavy ``string'' quasiparticles. In particular the autocorrelation function for $x \ll \sqrt{Dt}$, i.e., the ``return probability,'' vanishes with an anomalous exponent $\gamma = \min(\frac{1}{2} + \frac{|\mu|}{ \eta}, 1)$ throughout this phase. Generic response functions therefore behave anomalously at fixed $q$ when $\omega \rightarrow 0$. This behavior {is consistent with} extensive simulations using MPO methods (Fig.~\ref{MainFig}). At the isotropic point we wrote down a scaling form for the structure factor, and provided an elementary derivation of the dynamical critical exponent~\cite{gv_superdiffusion, denardis_superdiffusion}. Many possible extensions present themselves, including a systematic derivation of fluctuating hydrodynamics and long-time tails near half-filling, and an understanding of the scaling properties as one approaches $\Delta \rightarrow 1$ from the easy-plane (``gapless'') regime.

\emph{Acknowledgments}.---The authors thank Vincenzo Alba, Jacopo De Nardis, David Huse, Christoph Karrasch, and Vadim Oganesyan for helpful discussions. This work was supported by NSF Grant No. DMR-1653271 (S.G.), and US Department of Energy, Office of Science, Basic Energy Sciences, under Award No. DE-SC0019168 (R.V.). The authors are grateful to the KITP, which is supported by the National Science Foundation under Grant No. NSF PHY-1748958, and the Program ``The Dynamics of Quantum Information'', where part of this work was performed.

\bibliography{refs}

\bigskip

\onecolumngrid
\newpage



\section*{Supplementary material}

# Supplemental Material for "Anomalous relaxation and the spin structure factor of XXZ spin chains"

Sarang Gopalakrishnan,[1] Romain Vasseur,[2] and Brayden Ware[2, 3]

[1]*Department of Physics and Astronomy, CUNY College of Staten Island, Staten Island, NY 10314; Physics Program and Initiative for the Theoretical Sciences, The Graduate Center, CUNY, New York, NY 10016, USA*
[2]*Department of Physics, University of Massachusetts, Amherst, MA 01003, USA*
[3]*Rudolf Peierls Centre for Theoretical Physics, Clarendon Laboratory, University of Oxford, Oxford OX1 3PU, UK*

(Dated: April 1, 2019)

## I. INFINITE TEMPERATURE THERMODYNAMIC BETHE ANSATZ FORMULAS

We focus on the limit of infinite temperature, where closed form expressions for thermodynamic Bethe ansatz (TBA) formulas are available. The relevant expressions for us were computed in Ref. 1, but are summarized here for convenience. We focus on the Ising phase of the XXZ spin chain ($\Delta > 1$), and consider an infinite temperature state with chemical potential $\mu$ such that the local magnetization is given by $h = \langle \sigma^z \rangle = \tanh \mu$. The elementary scattering kernel in that regime reads

$$K_s(u) = \frac{1}{\pi} \frac{\sinh(\eta s)}{\cosh(\eta s) - \cos 2u}, \quad (1)$$

where $-\pi/2 \leq u \leq \pi/2$ is the rapidity, and $\Delta = \cosh \eta$. The occupation (Fermi factor) of the string $s$ is given by

$$\theta_s = \frac{\sinh^2 \mu}{\sinh^2(\mu(s+1))}, \quad (2)$$

so at half-filling, $\theta_s = 1/(s+1)^2$. Meanwhile, the dressed magnetization reads

$$m_s^{\rm dr} = \frac{1}{2} \frac{\sinh \mu(s+1)}{\sin \mu} \left( \frac{s}{\sinh s\mu} - \frac{s+2}{\sinh(s+2)\mu} \right), \quad (3)$$

which interpolates between the bare magnetization $m_s^{\rm dr} \sim s$ for heavy strings $s\mu \gg 1$, while $m^{\rm dr} \sim \frac{1}{3}(s+1)^2 \mu$ for light strings. Finally, the effective velocity of the strings is $v_s(u) = \varepsilon'(u)/(2\pi \rho_s^{\rm tot}(u))$ with the dressed energy is

$$\varepsilon'(u) = -\pi \sinh \eta \frac{\sinh(s+1)\mu}{\sinh 2\mu} \left( K'_s(u) \frac{\sinh \mu}{\sinh s\mu} - K'_{s+2}(u) \frac{\sinh \mu}{\sinh(s+2)\mu} \right), \quad (4)$$

and the total density of states is given by

$$\rho_s^{\rm tot}(u) = \frac{\sinh(s+1)\mu}{\sinh 2\mu} \left( K_s(u) \frac{\sinh \mu}{\sinh s\mu} - K_{s+2}(u) \frac{\sinh \mu}{\sinh(s+2)\mu} \right). \quad (5)$$

As expected from perturbation theory, the effective velocity scales as $v_s \sim e^{-(s-1)\eta}$ at large $\Delta$. These formulas are compatible give the following sum rule for the spin structure factor:

$$\int dx C(x,t) = \sum_{s=1}^{\infty} \theta_s(1-\theta_s)(m_s^{\rm dr})^2 \frac{\sinh \mu(1+s)}{2\cosh \mu} \left( \frac{1}{\sinh hs} - \frac{1}{\sinh h(s+2)} \right) = \frac{1}{4} \left( 1 - \tanh^2 \mu \right). \quad (6)$$

where we have used an explicit expression for $\int du \rho_s^{\rm tot}(u)$. This expression coincides with $\langle S_z^2 \rangle - \langle S_z \rangle^2$ as it should. The spin diffusion constant at half-filling is then given by [2] (see also Refs. 3, 4):

$$D = 2 \sum_s \int du \rho_s^{\rm tot}(u) \theta_s(1-\theta_s) \left| v_s(u) \right| \left. \frac{dm_s^{\rm dr}}{d\mu} \right|^2_{\mu=0} = \frac{2 \sinh \eta}{9\pi} \sum_{s=1}^{\infty} (1+s) \left[ \frac{s+2}{\sinh \eta s} - \frac{s}{\sinh \eta (s+2)} \right]. \quad (7)$$

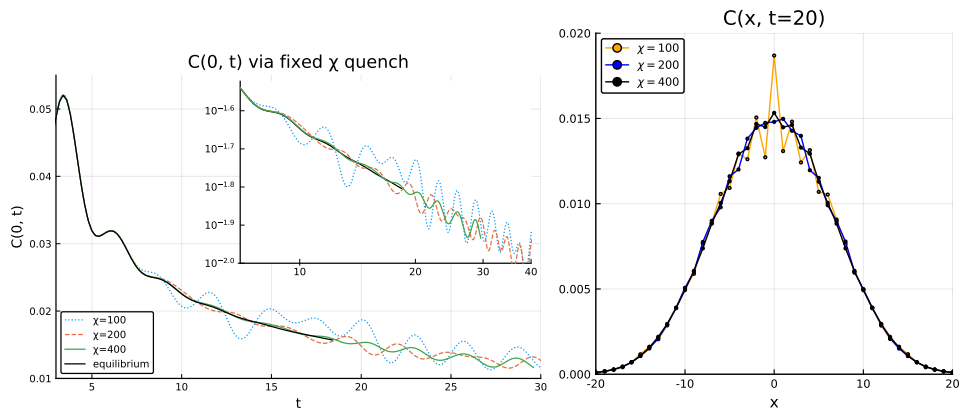

FIG. 1: **Convergence in the quench setup with fixed bond dimension.** (Left) Return probability $C(0,t)$ for $\eta = \frac{1}{2}$ at half-filling, computed from the quench setup with various fixed bond dimensions $\chi = 100, 200, 400$, compared to the equilibrium result computed using a fixed truncation error $\epsilon = 10^{-8}$ (reaching $\chi = 2000$ at the longest times). Inset: same plot in log-log scale. (Right) Convergence of the spatial profiles as a function of bond dimension. Even though the quench setup at fixed bond dimension can yield converged results for the full profile $C(x,t)$ [5, 6] — especially near the front, away from $x = 0$ — the return probability appears to be much more sensitive to truncation errors.

## II. QUENCH SETUP VS EQUILIBRIUM STRUCTURE FACTOR

We have explored two approaches for computing the structure factor $C(x,t) = \langle S_x^z(t) S_0^z(0) \rangle^c$. The first is to directly time-evolve the operator $S_z$ in the Heisenberg picture, as a matrix product operator (MPO). This approach corresponds precisely to performing the linear response calculation, so we refer to it as the equilibrium approach. In addition to being strictly in the linear response regime, the equilibrium approach has the advantage that one can study $\mu$-dependence systematically with minimal computational effort: since $C(x,t,\mu) = \text{Tr}[e^{-2\sum_i \mu S_i^z} S_x^z(t) S_0^z(0)]/Z - \langle S_z^2 \rangle$, the MPO $S_x^z(t)$ only needs to be computed once. In practice this computation can be further sped up [7] by instead evaluating $\text{Tr}[e^{-2\sum_i \mu S_i^z} S_x^z(t/2) S_0^z(-t/2)]$. These MPOs are only evolved for half as long and therefore have much lower bond dimension than $S_x^z(t)$.

Instead of this direct method, the structure factor can also be computed using a linear-response quantum quench. Following Ref. 6, we consider an initial density matrix $\rho(t=0) \propto (e^{(\mu+\delta\mu/2)\sigma_z})^{\otimes L/2} \otimes (e^{(\mu-\delta\mu/2)\sigma_z})^{\otimes L/2}$. The equilibrium (connected) spin structure factor can then be expressed as [6]

$$C(x,t) = \frac{1}{4} \lim_{\delta\mu \to 0} \frac{\langle \sigma_{x-1}^z(t) \rangle_{\text{quench}} - \langle \sigma_x^z(t) \rangle_{\text{quench}}}{\delta\mu}, \qquad (8)$$

where $\langle \ldots \rangle_{\text{quench}}$ refers to expectation values in this quench setting. We use MPO methods to time evolve the initial density matrix $\rho(t=0)$, and to evaluate expectation values of the spin. This provides another way to compute the spin structure factor. The quench setup could possibly allow us to reach longer times. While the bond dimension appears to grow more slowly in the quench setting, we find that very small truncation errors (less than $\epsilon = 10^{-10}$) are required to obtain converged results for the return probability ($x=0$) for $\delta\mu = 0.01$ (a value small enough to make eq. (8) correct up to negligible errors). On the other hand, as noted in Refs. 5, 6, working with fixed bond dimension $\chi$ in the quench setup appears to lead to relatively small errors for small values of $\eta$, yielding reasonably converged results for the full shape of $C(x,t)$ up to very long times $t \sim 200$ – especially near the front where truncation errors are less important. In contrast, the return probability $C(x=0,t)$ does seem sensitive to truncation errors, and even fixed $\chi = 400$ appears to deviate from the numerically exact equilibrium result at short times $t \sim 15$ (Fig 1). In this work, we thus use the quench setup with fixed bond dimension to compute the full profile of $C(x,t)$, as this yields reasonably converged results away from $x=0$ up to long times, and we restrict ourselves to fixed truncation error data computed directly in equilibrium for the return probability.

## III. DIFFUSIVE SCALING OF THE STRUCTURE FACTOR AT HALF-FILLING FOR $\Delta > 1$

For $\Delta > 1$, we expect spin diffusion at half-filling. Using the quench setup with fixed bond dimension $\chi = 400$, we computed numerically the full structure factor $C(x,t)$ for $\eta = \frac{1}{2}$ at half-filling. Its behavior is not perfectly Gaussian

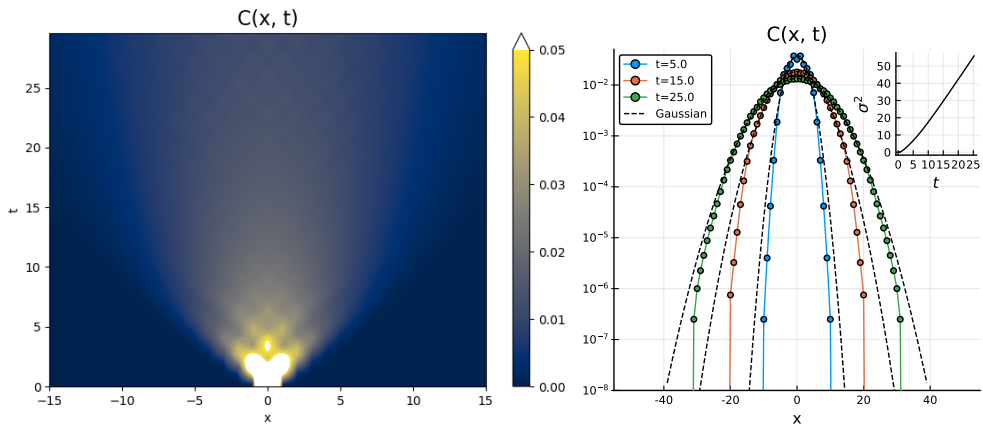

FIG. 2: **Structure factor profile for $\eta = 0.5$ and $\mu = 0$.** Left: spatio-temporal plot of the structure factor for $\eta = 0.5$, obtained from the quench setup with fixed bond dimension $\chi = 400$. Right: $C(x,t)$ for different times, and Gaussian fits. The tails of the structure factor are not exactly Gaussian for this value of $\eta$, probably due to the proximity to the isotropic point $\eta = 0$. Inset: the variance $\sigma^2(t) = 4\sum_x C(x,t)x^2$ shows the expected diffusive behavior at long times.

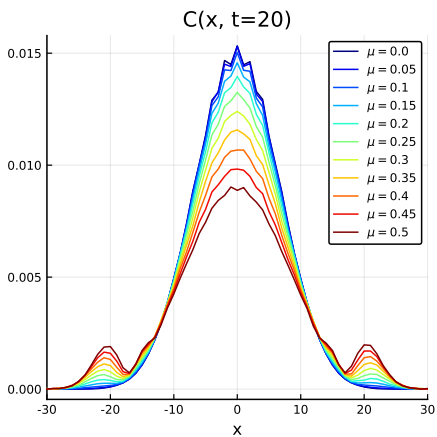

FIG. 3: **Spatial profiles for $\eta = 0.5$ away from half filling.** Spatial profiles at fixed time $t = 20$ for $\eta = 0.5$, obtained from the quench setup with fixed bond dimension $\chi = 400$. As $\mu$ is increased, the profiles show very clear ballistic features corresponding to light strings moving away from the origin ballistically.

on accessible time scales (Fig. 2), most likely due to the proximity to the isotropic point $\eta = 0$. For small $\eta$, we expect a crossover from superdiffusive to diffusive behavior on a time scale [2] $t_* \sim \eta^{-3}$. This is consistent with our data, as the variance $\sigma^2(t) = 4\sum_x C(x,t)x^2$ first grows superlinearly in time, and appears to then scale linearly at longer time scales, as expected from diffusion. This proximity to $\Delta = 1$ also explains why the return probability scales with an apparent exponent $\sim t^{-0.6}$ at half filling for $t < 20$, in between $\sim t^{-2/3}$ (expected for $t < t_*$) and $\sim t^{-1/2}$. As explained in the main text, working with larger $\eta$ further away from the isotropic point would make diffusion clearer, but our main prediction of continuously exponent is more easily testable at short times for small $\eta$. Away from half-filling, the spatial profiles develop clear ballistic features corresponding to light strings flying away ballistically (Fig. 3).

## IV. ISOTROPIC POINT $\Delta = 1$

The isotropic point exhibits spin superdiffusion [1, 5], with a time-dependent diffusion constant scaling as [2, 5, 6] $D(t) \sim t^{1/3}$. A very recent numerical work seems to suggest that this point is in the KPZ universality class [6]. This is consistent with the variance of the structure factor measured numerically, as it scales with time as $\sigma^2 \sim t^{4/3}$

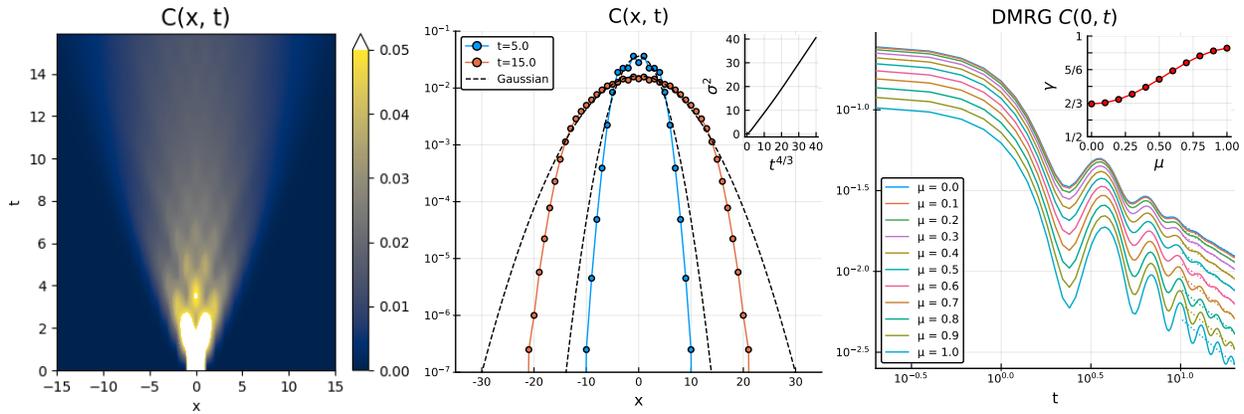

FIG. 4: **Structure factor at the isotropic point $\Delta = 1$ ($\eta = 0$).** Left: spatio-temporal plot of the structure factor at half-filling, obtained from the quench setup with fixed bond dimension $\chi = 200$. Middle: $C(x,t)$ for different times, and Gaussian fits. As noted in Ref. 6, the tails are evidently not Gaussian. Inset: anomalous scaling of the variance $\sigma^2(t) \sim t^{4/3}$. Right: Return probability $C(0,t) \sim t^{-\gamma}$ in the isotropic limit, for various chemical potentials, and power-law fits. Inset: Apparent exponent as a function of $\mu$.

(see Fig. 4). Figure 4 also shows the return probability $C(0,t)$ at the Heisenberg point $\Delta = 1$, for various chemical potentials $\mu$. As explained in the main text, we expect a crossover between $C(t) \sim t^{-2/3}$ for $t \ll \mu^{-3}$ (and in particular, at half-filling), and a ballistic behavior $t^{-1}$ at long times $t \gg \mu^{-3}$. Our DMRG data is consistent with $C(t) \sim t^{-2/3}$ at half-filling, but shows continuously evolving apparent exponents as a function of $\mu$. This illustrates the difficulty in distinguishing genuine continuously exponents as expected for $\Delta > 1$, from a crossover given the short times $t < 20$ accessible using DMRG.



\end{document}